# Decentralised Trust and Security Mechanisms for IoT Networks at the Edge: A Comprehensive Review

Khandoker Ashik Uz Zaman[1], Mahdi H. Miraz[1,2,3,*] and Mohammed N. M. Ali[1]

[1] Xiamen University Malaysia, Malaysia
[2] Wrexham University, UK
[3] University of South Wales, UK

## Abstract

INTRODUCTION: The proliferation of the amalgamation of IoT and edge computing has increased the demand for decentralised trust and security mechanisms capable of operating across heterogeneous and resource-limited devices. Approaches such as federated learning, Zero Trust architectures, lightweight blockchain and distributed neural models offer alternatives to centralised control.
OBJECTIVES: This review examines various state-of-the-art decentralised mechanisms and evaluates their effectiveness in terms of securing IoT networks at the edge.
METHODS: Thirty recent studies were analysed to compare how decentralised architectures establish trust, support secure communication and enable intrusion and anomaly detection. Frameworks, such as DFGL-LZTA, SecFedDNN and COSIER were assessed.
RESULTS: Decentralised designs enhance privacy, reduce single points of failure and improve adaptive threat response, though challenges remain in scalability, efficiency and interoperability.
CONCLUSION: The study identifies key considerations and future research needs for building secure and resilient trust-aware IoT edge ecosystems.

## 1. Introduction

The Internet of Things (IoT) has transformed into a global network of interconnected devices that autonomously sense, compute and exchange information across domains such as healthcare, smart transportation, industrial automation and urban infrastructure. This large-scale interconnectivity has accelerated the development of intelligent, data-driven systems while simultaneously creating significant challenges relating to trust, security, privacy and scalability. Centralised security architectures are traditionally dependent on perimeter-based defence and static trust assumptions. These are increasingly unsuitable for modern IoT deployments characterised by high device heterogeneity, resource constraints, continuous distributed data flows and the growing use of edge computing to support latency-sensitive operations [1,2].

In these decentralised and dynamic environments, establishing reliable trust between devices, services and users has become essential. IoT systems often lack central oversight and distributed operation creates vulnerabilities to identity spoofing, data manipulation, unauthorised device participation, insider attacks and poisoning-based disruptions [3]. Ensuring integrity, authenticity and trustworthy behaviour at the network edge, therefore, requires security mechanisms that operate autonomously, adaptively and independently of central authorities [4,5].

Amongst these mechanisms, several categories have gained prominence in recent literature. Recent research has introduced a range of decentralised trust and security mechanisms designed to address these challenges, including federated learning, distributed deep-learning models, Zero Trust Architecture (ZTA) and various decentralised trust-management frameworks.

Federated Learning (FL) enables collaborative training of models without centralising raw data, preserving privacy while distributing intelligence across edge devices. Frameworks such as SecFedDNN [6], Federated Trust Management [7] and collaborative intrusion detection systems [8,9] demonstrate improved detection accuracy and resilience against inference risks [10].

Deep Neural Networks (DNNs) contribute to decentralised anomaly detection, behavioural modelling and predictive analytics. When integrated with federated settings, they enhance robustness against manipulation and support adaptive security policies [11]. Examples include distributed DNN-based intrusion detection [12,13] and multi-layer learning for trust computation [14].

Zero Trust Architecture (ZTA) enforces continuous verification of every device and service, removing implicit trust even within internal networks [15]. Several recent studies apply ZTA within the context of decentralised IoT security, most notably DFGL-LZTA [16], which integrates federated graph learning and Zero Trust principles to strengthen edge security [17].

Decentralised architectural frameworks offer additional mechanisms for distributing trust by reducing reliance on central servers and enabling scalable authentication, policy enforcement and anomaly detection at the edge. Examples include COSIER [18], which decentralises consensus and trust computation across hierarchical clusters of devices and secure distributed identity frameworks [19,20] that shift verification processes to edge nodes. Several studies also propose multi-layer IoT trust systems [21,22] and edge-based trust architectures that utilise gateways or fog nodes for local decision-making and behavioural analysis. Such designs enhance resilience and responsiveness, especially in environments with intermittent connectivity or limited cloud access.

Although these approaches address complementary aspects of IoT trust and security, integrating them into fully decentralised ecosystems remains challenging. Federated learning systems face issues related to client heterogeneity, communication overhead, non-IID data distributions and vulnerability to poisoning attacks [23]. DNN-based models introduce computational complexity, require careful optimisation for resource-constrained devices and remain exposed to adversarial perturbations [24]. Zero Trust implementations demand fine-grained identity management and continuous policy enforcement, which are difficult to coordinate at scale [25]. In addition, decentralised trust frameworks often suffer from interoperability limitations, inconsistent trust metrics across heterogeneous platforms and a lack of standardised evaluation benchmarks for distributed IoT trust models [26].

These limitations highlight the need for a detailed review examining how decentralised mechanisms contribute to secure, trustworthy and scalable IoT operations at the edge. While earlier surveys have explored isolated technologies such as federated learning, anomaly detection or general IoT security models, comprehensive reviews that integrate decentralised trust, collaborative intelligence, behaviour-based security and Zero Trust edge architectures remain scarce and fragmented [27,28,29,30].

This review analyses thirty recent studies covering machine-learning-based trust models, distributed learning, decentralised access control, graph-based trust computation, Zero Trust policy enforcement and multi-layer IoT security architectures. It aims to:

(i) Classify current decentralised trust and security mechanisms for IoT edge systems.
(ii) Compare strengths, limitations and application domains of recent frameworks.
(iii) Identify open challenges and technological barriers.
(iv) Outline a forward-looking research roadmap for secure, adaptive and trustworthy IoT ecosystems.

## 2. Methodology

This review adopts a structured approach for identifying, screening and analysing recent research on decentralised trust and security mechanisms for IoT networks operating at the edge. The approach is highlighted in Figure 1. The objective was to ensure comprehensive coverage of developments in federated learning, machine-learning-based trust models, blockchain and lightweight blockchain frameworks [31], Zero Trust architectures, decentralised identity systems and hybrid edge security solutions.

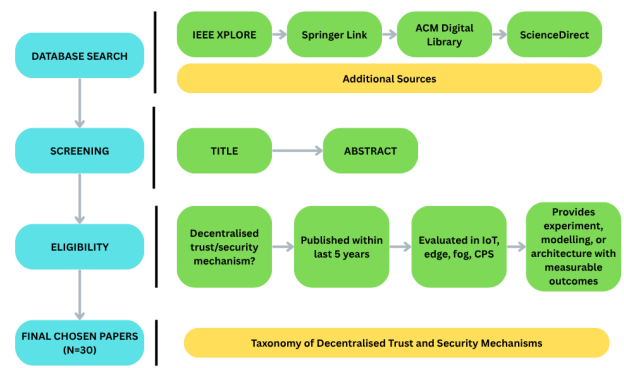

**Figure 1.** Review Paper Screening Methodology

## 2.1. Search Strategy

The initial literature search was performed across various major academic databases including IEEE Xplore, SpringerLink, ACM Digital Library and ScienceDirect. Additional studies were identified through backward and forward citation tracing. Search terms included IoT trust management, decentralised IoT security, federated learning intrusion detection, blockchain IoT framework (also known as Blockchain of Things (BCoT) framework, Zero Trust IoT, edge computing security and lightweight blockchain. Studies were included if they met the following criteria:

- The paper proposes, analyses or evaluates a decentralised trust, security or access-control mechanism for IoT systems.
- The work incorporates at least one decentralised technique: federated learning, deep learning, blockchain, lightweight blockchain, Zero Trust Architecture, graph learning or distributed identity management.
- The approach is designed for or evaluated within IoT, edge, fog or cyber-physical system environments.
- The study provides experimental evaluation, analytical modelling or an architectural framework with measurable outcomes.
- The publication date falls within the past five years, i.e. from 2021 until 2025, to ensure relevance to current research trends.

## 2.2. Paper Screening Process

A multi-stage screening procedure was applied:

(i) Title and abstract screening to remove clearly irrelevant works.
(ii) Full-text assessment to confirm technical relevance and alignment with the scope of decentralised IoT trust and security.
(iii) Thematic categorisation of the remaining studies into major areas, including:

- Federated learning-based intrusion and anomaly detection systems
- Deep-learning-based trust modelling and behavioural analysis
- Blockchain and lightweight blockchain trust frameworks
- Zero Trust IoT models
- Decentralised identity, reputation and access-control mechanisms; and
- Hybrid or graph-enhanced security architectures

A total of thirty papers met all inclusion criteria. These selected works formed the basis of this literature review, including the comparative analysis presented in Sections 4 and 5.

## 3. Background and Review of Decentralised IoT Trust Mechanisms

### 3.1. Trust Management in IoT

Trust management is a foundational requirement in IoT systems, where devices, services and users interact autonomously without guaranteed central oversight. Trust-related decisions typically consider three categories: identity trust, which verifies whether a device or a user is genuine; data trust, which evaluates the integrity and reliability of exchanged information; and behavioural trust, which assesses historical actions to identify anomalies or malicious patterns, as widely discussed in recent trust-management surveys [27].

Behaviour-based and reputation-based trust systems form a major subset of existing works. These approaches maintain local or distributed records of device interactions and dynamically update trust scores to reflect changing behaviour. DeepTrust, proposed by Ullah *et al.* [32], applies deep neural networks to interaction logs and contextual data to generate adaptive trust and reputation values. NeuroTrust, introduced by Awan *et al.* [33], employs multilayer perceptron models to calculate trust values in enormous Internet of Medical Things (IoMT) networks by integrating direct trust calculations, such as trustworthiness parameters like reliability and packet delivery ratio, with the aid of indirect clues to counter attacks such as the on-off attack, the whitewashing attack, the good-mouthing attack and the bad-mouthing attack.

A related group of studies focuses on anomaly and intrusion detection as a proxy for trust evaluation. For example, Douiba *et al.* [9] demonstrated that CatBoost-based gradient boosting can achieve high-accuracy anomaly detection across multiple IoT traffic datasets. The BFLIDS framework by Begum *et al.* [14] integrates federated learning and blockchain to train CNN/BiLSTM intrusion detectors in a decentralised and privacy-preserving manner. Although these approaches do not compute explicit trust scores, they contribute to trust management by estimating the likelihood of malicious behaviour.

Identity and credential management in decentralised environments has been addressed more directly in BETAC-IoT by Odeh & Taleb [19], which combines blockchain, federated learning, smart contracts and Merkle-tree verification to manage decentralised device identities and access policies. Other blockchain-enabled and FL-enabled frameworks, such as the reinforcement-learning-based system by Mohammed *et al.* [22], prioritised secure processing, reduced latency and adaptive detection in fog cloud environments rather than explicit reputation modelling.

Several reviewed architectures employ multi-layer or hierarchical trust-evaluation structures. NeuroTrust [33] uses hybrid servers on patient and provider sides for trust computation in IoMT networks. El-Sofany *et al.* [21] integrated ML-based intrusion detection with SDN/NFV to form a layered security architecture. SecFedDNN, proposed by Alamir et al. [6], distributes federated deep-learning-based intrusion detection across device, edge and cloud tiers using layer-adaptive sparsified aggregation. While not all of these systems explicitly maintain trust scores, they collectively support decentralised trust management by distributing behavioural analysis and update verification across the IoT edge ecosystems.

### 3.2. Decentralisation in IoT Security

Decentralisation is increasingly necessary in IoT systems because centralised architectures struggle with real-time constraints, bandwidth limitations and reliability demands in large, heterogeneous environments [1,2,3]. Central servers create bottlenecks, increase latency and concentrate security risk. By shifting computation and decision-making closer to devices, edge and fog computing allow systems to process data with greater context-awareness and reduced dependency on cloud resources located far away.

Decentralised security frameworks distribute authentication, trust scoring, anomaly detection and policy enforcement across multiple nodes. In hierarchical architectures, constrained IoT devices handle lightweight monitoring while fog or edge nodes perform in-depth analysis. This hierarchical multi-layer security model appears in multi-layer ML-based designs such as those by El-Sofany et al. [21] and in hybrid WSN–IoT intrusion detection systems developed by Karthikeyan et al. [34]. Asha et al.'s smart-city optimisation model [7] similarly demonstrates how processing tasks can be offloaded to distributed layers to improve scalability and reduce central reliance.

Blockchain-oriented decentralisation provides additional benefits by eliminating single points of failure. Mershad's COSIER system [18] proposes grouping IoT devices into clusters, maintaining lightweight local chains and periodically anchor cryptographic summaries to a global blockchain for integrity verification. This design follows a hierarchical anchoring model rather than a sharding-based architecture, as transaction execution and validation remain local and are not processed in parallel across shards. GTxChain by Cai et al. architecture [8] adopts a DAG-based blockchain with graph neural networks to support decentralised validation in high-volume IoT deployments.

Decentralised intelligence also appears in federated and hybrid learning systems. Surveys and models by Ferrag et al. [20], Mohammed et al. [22] and Nazir et al. [35] show that intrusion detection and anomaly analysis can be trained collaboratively across distributed nodes without sharing raw data. Reinforcement-learning-driven approaches such as MORFLB [22] and asynchronous FL techniques developed by Manu et al. [13] demonstrate that local nodes can refine detection thresholds and update policies without cloud synchronisation.

Decentralisation offers several advantages such as improved resilience, reduced latency and elimination of central points of failure. However, it also introduces challenges, including maintaining consistent trust states, synchronising model updates across heterogeneous environments and enforcing distributed access-control decisions without relying on a central authority [20,17,22].

### 3.3. Federated Learning in IoT

Federated Learning (FL) enables IoT devices to collaboratively train machine-learning models without transferring raw data, making it highly suitable for privacy-sensitive and bandwidth-constrained environments. Local devices compute training updates and send model gradients or parameters to an aggregator, preserving data locality while enabling global learning.

Many studies, covered in this review, applied FL to intrusion detection, anomaly detection and behaviour modelling. Asha et al. [7], Cai et al. [8] and Douiba et al. [9] used FL to detect malicious traffic and compromised devices by training classifiers on local behavioural indicators and system logs. Alamir et al.'s SecFedDNN [6] integrated FL with deep neural networks to improve distributed intrusion detection accuracy. Similar approaches appeared in Begum et al. [14], Nazir et al. [35], Alzubi et al. [36] and Manu et al. [13], covering healthcare, vehicular IoT and industrial systems, demonstrating that FL can achieve high accuracy across diverse IoT domains.

Hybrid architectures combine FL with blockchain, decentralised access control or optimisation techniques. Zhou et al.'s DFGL-LZTA [16] integrates federated graph learning with Zero Trust principles for continuous client validation. Odeh & Taleb's BETAC-IoT [19], Mohammed et al. [22], Gugueoth et al. [37] and Javed et al. [38] similarly explored combinations of FL with blockchain, smart contracts or optimisation models to improve security, scalability and trustworthiness.

Despite these advantages, several challenges remain significant in IoT-based FL. Devices frequently generate non-IID data, leading to unstable convergence in global models [39]. Client heterogeneity in computational power, battery capacity and connectivity results in inconsistent update rates and fairness issues [20]. FL is also vulnerable to poisoning and backdoor attacks, where malicious clients submit crafted gradients to corrupt the shared model; mitigation strategies such as client filtering, anomaly-based gradient checks and robust aggregation are examined by Ferrag et al. [20], Ullah et al. [32], Nazir et al. [35] and Latif et al. [39]. Furthermore, communication overhead remains a critical limitation, particularly for deep models, and has prompted techniques such as asynchronous updates [13], gradient compression and selective client participation.

Comprehensive surveys by Sah et al. [10], Ferrag et al. [20], Dritsas et al. [30] and Jiang et al. [40] highlighted the absence of standard evaluation benchmarks, consistent performance metrics and secure aggregation protocols across existing FL-based IoT systems. Overall, the reviewed studies show that FL offers strong privacy protection and supports collaborative learning at the edge. However, ensuring secure, scalable and trustworthy FL requires addressing data heterogeneity, improving update integrity, reducing communication cost and integrating FL with complementary decentralised trust mechanisms.

### 3.4. Deep Neural Networks for Decentralised Detection

Deep neural networks (DNNs) have become central to modern IoT security because they can learn complex, non-linear patterns in traffic and behavioural data, enabling accurate detection of evolving threats that static rule-based systems struggle to identify [11]. In intrusion detection, DNN-based systems including CNNs, RNNs and hybrid models have demonstrated strong performance across datasets such as NSL-KDD, UNSW-NB15, CICIDS2017 and Bot-IoT.

Alsubaei *et al.* [41] presents a smart deep-learning model combining extensive preprocessing, XGBoost and a tuned sequential neural network to classify multi-class intrusions with high accuracy. El-Sofany *et al.* [21] compares multiple machine-learning classifiers for IoT and shows that ML-driven IDS can autonomously handle evolving threats. Karthikeyan *et al.* [34] employs Firefly-optimised feature selection and SVM tuning in WSN–IoT environments to reduce computational load while improving detection performance. Gugueoth *et al.*'s survey [37] highlights widespread adoption of deep-learning-based IDS in decentralised IoT systems.

DNNs are also used in trust-scoring frameworks. NeuroTrust [33] applies an ANN to evaluate trust scores based on latency, packet loss and behavioural metrics. Deep Trust [32] updates reputation dynamically using deep models that adapt to changing attack patterns and misbehaviour.

DNNs increasingly appear in decentralised architectures. SecFedDNN [6] trains deep models using federated learning across device, edge and cloud tiers. Nazir *et al.* [35] integrates DNNs with blockchain-supported FL for secure update aggregation in IoT. Begum *et al.* [14] combines blockchain, FL and DNNs for IoMT intrusion detection. Manu *et al.* [13] demonstrates that asynchronous FL with deep models can reduce latency and improve practicality for edge deployments. Ferrag *et al.* [20] experimentally evaluates federated deep learning across real IoT datasets including Bot-IoT, MQTTset and TON_IoT.

However, DNN-based security introduces significant challenges. Deep models can overload constrained IoT devices due to their computational and memory requirements. They are vulnerable to adversarial examples, gradient poisoning and backdoor attacks, which can degrade or manipulate detection accuracy. Their lack of interpretability complicates forensic analysis and transparency in trust-related decisions. Recent surveys by Latif *et al.* [39] and Gugueoth *et al.* [37] emphasise the importance of combining DNNs with robust aggregation, malicious-client filtering, feature optimisation and privacy-preserving techniques to ensure secure deployment in decentralised IoT systems.

### 3.5. Zero Trust Architecture (ZTA) in IoT

Zero Trust Architecture (ZTA) eliminates assumptions of inherent trust within networks by enforcing continuous verification, least-privilege access and adaptive, context-aware authorisation. This paradigm is highly applicable to IoT, where devices frequently join and leave networks, operate autonomously in untrusted environments and communicate across multiple domains [17,25].

A key example of ZTA integration in decentralised IoT systems is DFGL-LZTA [16], which embeds a lightweight ZTA layer within a federated learning model. This framework incorporates continuous authentication, reputation-guided aggregation, local differential privacy and homomorphic encryption to protect model updates. A PPO-based deep reinforcement learning agent dynamically adjusts trust and access decisions based on behavioural and contextual factors.

Other works apply ZTA principles in different IoT contexts. Mohammed *et al.* [22] integrates continuous verification in federated fog cloud systems, updating trust and access permissions based on contribution quality and behaviour. Khan *et al.* [42] proposes a context-aware ZTA approach for connected and autonomous vehicles, where decisions involving routing, sensor access and inter-vehicle communication are evaluated using behavioural analytics and location data. Tanque *et al.* [25] emphasises micro-segmentation and strict identity validation for IoT platforms, while Denzel [28] highlights challenges in policy orchestration, identity management and cross-domain trust propagation in distributed IoT networks.

When combined with federated learning and machine-learning-based detection mechanisms, ZTA enables real-time policy enforcement that adapts as device behaviour evolves. This ensures that trust is not granted permanently but recalculated continuously based on updated evidence.

When combined with federated learning and machine-learning-based detection, ZTA supports real-time, adaptive security enforcement. However, practical deployment remains challenging due to continuous monitoring overhead, identity-management complexity and the difficulty of implementing fine-grained policies on resource-constrained devices. Lightweight ZTA variants, such as those in DFGL-LZTA [16], are therefore essential for realistic deployment at the network edge.

### 3.6. Edge Computing as a Foundation for Decentralised Trust

Edge computing plays a central role in decentralised IoT security because it brings computation, storage and decision-making closer to devices. This reduces reliance on cloud servers, lowers latency and enables local enforcement of trust and security policies. In distributed IoT environments, edge nodes frequently perform tasks, such as data preprocessing, behaviour analysis, anomaly detection, access control and trust evaluation, making edge computing an essential foundation for decentralised architectures [1,2].

Many decentralised systems rely on edge nodes to execute security-critical functions such as traffic analysis, intrusion detection and trust evaluation. Machine-learning-based intrusion detection frameworks, including those proposed by El-Sofany et al. [21] and Alsubaei et al. [41], deployed classifiers at the edge to allow real-time threat detection without transmitting raw traffic to cloud environments. Similarly, feature-selection-based and reinforcement-learning-driven IDS systems for WSN–IoT networks offload computation to fog or edge layers, reducing the burden on constrained sensor nodes [34].

Edge computing also enables decentralised learning methodologies. Federated learning frameworks such as SecFedDNN [6], asynchronous FL [13], BFLIDS [14] and blockchain-assisted FL [35] rely on edge devices or gateways to train local models, validate gradients, filter malicious updates and communicate with aggregators. This arrangement preserves data privacy by keeping raw data local and mitigates the overhead associated with transmitting large datasets to cloud servers [10,20,30].

Several hybrid frameworks integrated edge computing with blockchain or distributed ledger technologies. The COSIER architecture by Mershad [15] places lightweight blockchain components at edge clusters, where local leaders maintain cluster-level chains and periodically synchronise with a global ledger. BETAC-IoT by Odeh & Taleb [19] employed hybrid public–private blockchains at the edge to support scalable identity management and access control. Nazir et al. [35] extends this concept by integrating blockchain-backed aggregation into an FL–DNN system and quantifying trust improvements through metrics, such as the Security Efficacy Metric.

Edge computing also supports context-aware trust decisions. Khan et al.'s zero-trust design for autonomous vehicles assigns verification, sensor validation and behaviour monitoring to edge-level components before data transmission to higher layers [42]. Mohammed et al. [22] uses a fog cloud architecture where reinforcement learning, and trust scoring occur at the edge to support secure transport scheduling. These examples show that the edge layer is crucial for rapid, context-sensitive trust assessment.

Across these systems, edge computing consistently serves as the operational layer where decentralised trust mechanisms are enforced. However, deploying advanced models at the edge remains challenging due to computational limits, memory constraints, hardware diversity and intermittent connectivity. Lightweight designs such as COSIER [18] and DFGL-LZTA [16] demonstrate that decentralised trust solutions must be tailored to edge capabilities to remain practical at scale.

## 3.7. Blockchain and Distributed Ledger Systems for IoT Trust

Blockchain and distributed ledger technologies (DLTs) have emerged as key enablers of decentralised trust in IoT networks. By providing immutable and tamper-evident records without the need for a central authority, blockchain allows devices to verify trust scores, identity credentials, policy updates and transactions in a transparent and secure manner. This is particularly valuable in IoT environments where devices operate autonomously and may not inherently trust one another [43].

D'Aniello et al.'s survey provides a comprehensive overview of blockchain-based trust systems, examining how smart contracts, reputation mechanisms and blockchain-backed identity registries support secure interactions. The review notes that blockchain can not only store trust values, log device behaviour, enforce access-control policies and coordinate distributed decision-making, but also identifies challenges including latency, scalability issues and energy consumption [43].

Lightweight blockchain frameworks address these constraints by optimising cryptography, consensus mechanisms and ledger structure. Several studies demonstrate that adapting blockchain design is essential for IoT environments characterised by limited computation and energy. Mershad's COSIER architecture [18], for example, organises IoT devices into clusters with lightweight local ledgers coordinated through a higher-level verification mechanism, reducing communication and processing overhead.

Other frameworks integrate blockchain with federated learning and ML-based detection. Nazir et al. [35] combined FL, DNNs and blockchain to secure model updates and introduce metrics, such as the Security Efficacy Metric and Comparative Improvement Factor. BETAC-IoT [19] used hybrid blockchain, smart contracts and Merkle-tree verification to support decentralised identity management at the edge. Alzubi et al. [36] integrates blockchain with FL and CNN-based analysis for secure handling of electronic health records in Cloud–IIoT environments. Mohammed et al. [22] and Ferrag et al. [20] similarly applied blockchain to reinforce tamper resistance, auditability and secure coordination in fog–cloud or FL-driven IoT systems.

Overall, blockchain serves as a distributed backbone for trust establishment, identity verification, secure model exchange and behaviour logging. Nevertheless, widespread adoption remains constrained by computational cost, energy consumption and communication overhead, reinforcing the need for lightweight and hybrid ledger designs tailored to IoT environments.

## 3.8. Decentralised Access Control and Identity Management

Access control governs which devices, users or applications are permitted to perform specific operations within an IoT system. Traditional access-control models rely on central authorities to enforce role-based or attribute-based rules, but such centralisation becomes impractical in large, heterogeneous and geographically distributed IoT networks. As IoT ecosystems move toward edge-driven architectures, access control must become decentralised, adaptive and context-aware.

Several studies demonstrate how decentralised access control can be achieved by distributing authorisation logic across edge nodes rather than relying on a single authentication server. Blockchain-based designs commonly employ smart contracts to automate authorisation decisions using stored identity credentials, trust information and policy rules, thereby reducing reliance on centralised control points and lowering authentication latency [19]. These mechanisms enable consistent policy enforcement in distributed environments while supporting scalability.

Other designs adopt semi-decentralised or fully decentralised access-control models. Mohammed *et al.*'s fog cloud architecture [22] updates trust and access permissions dynamically via reinforcement learning operating at the edge, turning gateways into local decision-making nodes. Khan *et al.*'s context aware Zero Trust approach for connected vehicles [42] bases access decisions on behavioural monitoring, sensor validation and location context rather than static roles or credentials.

Decentralised identity management is also highlighted in several studies. Ferrag *et al.* [20] examined distributed ledgers for secure device registries, cryptographic identifiers and tamper-resistant trust records. Alzubi *et al.* [36] integrated ledger-based identity verification into FL-driven healthcare applications to secure electronic health records. In many of these systems, ML or anomaly detection modules automatically downgrade or revoke identities when suspicious activity is detected.

Overall, the literature reveals a transition towards trust-integrated, behaviour-aware and decentralised access control. Rather than relying solely on static certificates, modern IoT systems combine identity data, trust scores, behavioural analytics and blockchain-backed registries to enforce adaptive and fine-grained authorisation in distributed environments.

## 3.9. Graph-Based, DRL-Enhanced and Hybrid Security Models

Advanced IoT security frameworks increasingly integrate graph learning, deep reinforcement learning (DRL) and hybrid trust mechanisms to model complex device relationships and adapt to evolving threats. These approaches aim to capture structural interactions between IoT nodes, dynamically adjust security policies and combine multiple decentralised techniques into unified defence systems.

A leading example is the DFGL-LZTA framework [16], which combines decentralised federated graph learning with a lightweight Zero Trust Architecture. Device communication patterns are represented as graphs, and a hierarchical Graph Attention Network (GAT) is trained locally to identify community structure, anomalous edges and high-risk nodes. Sparse global aggregation reduces communication overhead by sharing only essential graph embeddings. A PPO-based DRL agent dynamically updates Zero Trust policies based on contextual and behavioural data, while local differential privacy and homomorphic encryption secure model updates [16].

Other studies integrate graph-based reasoning or DRL for adaptive trust decisions. NeuroTrust [33] models trustworthiness using behavioural indicators across interaction networks, while Deep Trust [32] extends decentralised reputation management using deep-learning-driven behavioural analysis. Reinforcement learning also appears in fog–cloud transport systems, where Mohammed *et al.* [22] employed DRL to optimise resource allocation and trust decisions dynamically. Baksh *et al.* [44] explores DNN-based intrusion detection that can be augmented with adaptive strategies driven by contextual feedback.

These hybrid models offer fine-grained, context-aware security by combining graph topology, behaviour modelling and dynamic policy optimisation. However, they also introduce computational complexity, training instability and reduced interpretability when deployed in decentralised environments. Balancing accuracy, efficiency and transparency remains an active challenge for next-generation graph- and DRL-enhanced IoT security systems.

## 3.10. Multi-Layer and Distributed IoT Security Architectures

Multi-layer and distributed architectures are widely adopted to address scalability, heterogeneity and real-time detection requirements in IoT systems. These architectures distribute computation and decision-making across edge, fog and cloud layers, avoiding bottlenecks and eliminating single points of failure.

Several studies follow an edge–fog–cloud tiered structure. Manu *et al.* [13] proposes an intelligent edge security architecture where lightweight anomaly filtering occurs at the edge while deeper classification occurs at higher layers. Sumathi *et al.* [12] introduces a layered IDS in which packet sampling and preprocessing occur at edge routers while classifiers operate at upper tiers to detect attacks such as DDoS and spoofing.

Other frameworks add dedicated trust or identity layers. Mershad [18] defines separate layers for identity verification, trust scoring and service access in fog-supported authentication. NeuroTrust [33] employs hierarchical trust computation in IoMT networks, with trust aggregated upwards for global reliability assessment. Deep Trust [32] distributes behaviour monitoring and reputation updates across layers to improve scalability.

Blockchain-augmented multi-layer designs also appeared in our reviewed literature. Ferrag *et al.* [20] describes a multi-level blockchain-supported architecture where identities, logs and trust scores are stored across distributed ledger units. BETAC-IoT [19] combines identity registration, smart-contract evaluation, federated learning and access control into coordinated layers.

Hybrid multi-layer systems are also used in smart transport and smart-city environments. Mohammed *et al.* [22] integrates DRL-based decision-making across fog and cloud layers, while works by Alsubaei *et al.* [41], El-Sofany *et al.* [21] and Gugueoth *et al.* [37] adopt layered ML detection pipelines to secure healthcare and smart-home networks.

These architectures reduce latency, improve scalability and increase fault tolerance by distributing the security responsibilities. However, coordinating policies and ensuring consistent trust decisions across layers remains a complex task. This is particularly true in dynamic IoT environments [39].

## 3.11. Privacy-Preserving and Lightweight Cryptographic Techniques in IoT

Privacy-preserving techniques are essential in decentralised IoT systems because devices routinely process sensitive information while operating under strict computational and energy constraints. Several studies, therefore, introduced mechanisms that protect data confidentiality, secure model updates and safeguard device identity without relying on a central authority.

A prominent category of approaches involves privacy-preserving federated learning, where raw data never leaves edge devices. The DFGL-LZTA framework applies local differential privacy to mask model updates prior to transmission and incorporates homomorphic encryption to secure gradients during aggregation [16]. These measures ensure that private information cannot be reconstructed even if communication channels or aggregators are compromised. Similar objectives appear in collaborative DNN–FL–blockchain architectures, which emphasise privacy protection for sensitive data in healthcare, industrial and vehicular environments [35].

Lightweight cryptographic techniques represent another critical theme, particularly for constrained IoT nodes that cannot support conventional cryptographic algorithms. The COSIER architecture incorporates ASCON-a NIST-selected lightweight authenticated encryption scheme, together with simplified consensus mechanisms and compact ledger structures to minimise latency, memory usage and energy consumption [18]. Experimental evaluations in COSIER show reduced transaction delay, lower communication overhead and improved energy efficiency compared with conventional blockchain designs, indicating that lightweight cryptographic protection can be deployed within the resource limits of edge devices.

Other systems employ Merkle-tree-based integrity verification and blockchain-backed identity management to enhance privacy and tamper resistance. The BETAC-IoT framework uses a hybrid blockchain and Merkle-tree structure to verify data integrity and regulate access control without exposing raw identity information or behavioural logs [19]. Privacy-aware trust frameworks such as Deep Trust similarly implement secure reputation propagation, allowing nodes to exchange trust updates without disclosing sensitive interaction histories [32].

Collectively, these techniques illustrate a broader shift toward privacy-preserving decentralised security, where lightweight encryption, secure aggregation, privacy masking and efficient ledger designs support trustworthy operation across heterogeneous IoT environments. Remaining challenges include balancing strong privacy guarantees with resource limitations at the edge, minimising cryptographic overhead and ensuring that encrypted or obfuscated data remain sufficiently informative for accurate model training and trust evaluation.

## 4. Taxonomy of Decentralised IoT Trust and Security Approaches

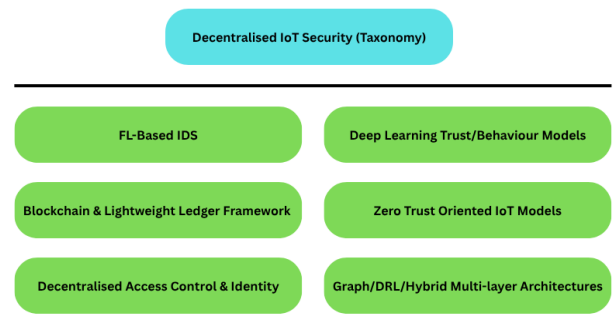

**Figure 2.** Taxonomy of Decentralised Trust and Security Mechanisms

The thirty selected studies can be organised into six primary categories according to their core decentralised trust or security mechanisms, as shown in figure 2. These categories reflect the dominant techniques, architectural patterns and threat models addressed in recent IoT–edge security research. The taxonomy highlights how different approaches contribute complementary functions, ranging from decentralised learning and behaviour monitoring to tamper-proof coordination and adaptive access control.

## 4.1. Federated Learning–Based Intrusion and Anomaly Detection

A substantial portion of the reviewed research employs federated learning to perform intrusion detection and anomaly analysis across distributed IoT, edge or fog environments. These systems train local models on device-resident data and aggregate updates without exposing raw information. Representative frameworks include SecFedDNN [6], BFLIDS [14], Cai *et al.* [8], Douiba *et al.* [9], Manu *et al.* [13], Nazir *et al.* [35], Ferrag *et al.* [20] and Alzubi *et al.* [36]. Techniques found in this category include CNN/BiLSTM hybrids, asynchronous FL, gradient compression, selective participation and blockchain-secured update aggregation.

These approaches improve privacy, reduce reliance on central servers and support collaborative defence against evolving attacks. However, they also confront challenges including non-IID data distributions, communication overhead and vulnerability to poisoning attacks—these issues have been analysed in detail across several studies.

## 4.2. Deep-Learning-Based Trust and Behavioural Analysis

Another set of works applies deep neural networks to trust evaluation and behavioural modelling. These systems infer trust scores or detect malicious activity based on learned features from traffic, contextual attributes or historical interaction data. Trust-specific DNN frameworks include Deep Trust [32] and NeuroTrust [33], while other studies apply deep or optimised neural networks for traffic classification or anomaly detection [21,34,41].

This category emphasises continuous, data-driven assessment of device behaviour and enables rapid adaptation to threat that change over time. Although highly effective, these models can be computationally demanding and may require lightweight adaptations for deployment on constrained IoT nodes.

## 4.3. Blockchain and Lightweight Distributed Ledger Frameworks

Blockchain and DLT-based systems support decentralised trust establishment by providing immutable audit trails, distributed identity management and tamper-resistant update verification. Lightweight designs such as COSIER [18] reduce computation and storage overhead by using simplified consensus, cluster-level chains and compact block formats. Hybrid ledger–FL frameworks, including BETAC-IoT [19], Nazir *et al.* [35], Alzubi *et al.* [36] secure model updates, store trust records and automate access-control decisions via smart contracts.

These frameworks enhance transparency and resist tampering but also introduce challenges relating to energy consumption, block-generation delay and network overhead. Lightweight and hybrid ledger structures address these constraints by adapting blockchain for resource-limited environments.

## 4.4. Zero Trust–Based IoT Security Models

Zero Trust architectures eliminate implicit trust and enforce continuous authentication, micro-segmentation and context-aware authorisation. DFGL-LZTA [16] is a key representative, integrating Zero Trust principles with federated graph learning and incorporating reinforcement learning for dynamic policy adaptation. Other ZTA-oriented works include Mohammed *et al.* [22], Khan *et al.* [42], Tanque *et al.* [25] and Denzel [28] which apply continuous verification in domains ranging from autonomous vehicles to fog-cloud transport systems.

These models strengthen device authentication, identity validation and policy enforcement in highly dynamic IoT ecosystems. Implementation challenges include computational overhead and the need to manage frequent authentication checks on constrained nodes.

## 4.5. Decentralised Access Control and Identity Management

Decentralised access-control frameworks distribute credential verification, authorisation logic and policy updates across edge nodes to reduce reliance on central authorities. BETAC-IoT [19] is a leading example, combining blockchain, federated learning and Merkle-tree verification to implement trust-aware access control. Additional contributions appear in Mohammed *et al.* [22], Khan *et al.* [42], Ferrag *et al.* [20] and Alzubi *et al.* [36] which integrate behaviour-based trust assessment, distributed identity validation or anomaly-driven credential revocation.

These models enable fine-grained, real-time access decisions based on contextual risk rather than static roles or credentials. They also support scalability by allowing gateways or edge nodes to act as local policy decision points.

## 4.6. Hybrid Graph-Based, DRL-Enhanced and Multi-Layer Architectures

A final category includes advanced hybrid models that combine graph learning, deep reinforcement learning, hierarchical DNNs or multi-layer intrusion detection pipelines. DFGL-LZTA [16] again appears here due to its integration of graph attention networks and reinforcement learning within a decentralised Zero Trust framework. Other hybrid approaches incorporate deep-learning-based trust computation [32] hierarchical trust aggregation [33], reinforcement-learning-based optimisation [22], and layered IDS architectures [12,34]. Additional contributions include smart-city and fog–cloud security systems that blend multiple decentralised mechanisms [7, 37].

These models offer adaptive, context-aware security but also introduce additional computational complexity and require careful design to remain practical in constrained IoT environments.

## 5. Comparative Analysis of Reviewed Frameworks

This section provides a comparative evaluation of the thirty reviewed frameworks across several key dimensions, using the taxonomy counts in Table 1 and the qualitative design assessments in Table 2 and Table 3 as reference points. Table 1 summarises the distribution of studies across the major architectural categories identified in Section 4, including federated-learning-driven intrusion detection, blockchain-centric trust frameworks, Zero Trust–based security models and decentralised access-control mechanisms. This distribution illustrates the prominence of specific techniques, particularly federated learning and deep-learning-based detection in recent decentralised IoT security research.

Table 2 evaluates each framework against three core criteria: performance, scalability and robustness. These scores reflect an independent assessment based on reported experimental results, architectural design assumptions and the operational context described in the original studies. Performance considerations include intrusion detection accuracy, latency and communication overhead; scalability relates to the ability to support large or heterogeneous device populations; and robustness examines resilience against adversarial attacks, poisoning, spoofing or model manipulation.

Table 3 complements this assessment by comparing the frameworks in terms of privacy preservation, degree of decentralisation and practicality for resource-constrained devices. Privacy assessments focus on whether raw data remain local, whether model updates are protected cryptographically and how identity information is managed. Decentralisation strength considers the extent to which decision-making, trust evaluation and model aggregation are distributed across edge or fog nodes. Practicality evaluates computational cost, memory requirements and feasibility of deployment on low-power IoT devices.

Together, these tables and the accompanying taxonomy illustrated in Figure 2 provide a holistic view of how recent decentralised IoT security frameworks differ in design goals, underlying mechanisms and deployment constraints. The comparative analysis helps identify recurring strengths such as improved privacy and resilience in federated and blockchain-based designs as well as common limitations, including communication overhead, susceptibility to poisoning attacks and reliance on high-complexity models that may not suit resource-constrained edge devices.

Table 1. Categories of reviewed frameworks

| Category | No. of Reviewed Papers |
|---|---|
| Federated learning-based IDS/anomaly detection | 12 |
| Deep-learning-based trust/behavioural models | 10 |
| Blockchain & lightweight ledger frameworks | 6 |
| Zero trust oriented IOT models | 6 |
| Decentralised access control & identity | 7 |
| Graph/DRL/hybrid multi-layer architectures | 9 |

Table 2. Qualitative scoring for each category of reviewed frameworks according to computational cost, robustness and scalability

| Framework Category | Performance Metrics | | |
|---|---|---|---|
| | Comp. Cost | Robustness | Scalability |
| Federated Learning Based IDS/Anomaly Detection | Medium | Medium | High |
| Deep-learning-based trust/behavioural models | High | Medium | Medium |
| Blockchain & Lightweight Ledger Framework | Medium | High | Medium |
| Zero Trust Oriented IOT Models | Medium | High | Medium |
| Decentralised Access Control & Identity | Low | Medium | High |
| Graph / DRL / Hybrid Multi-layer Architectures | High | High | Medium |

Table 3. Qualitative scoring for each category of reviewed frameworks according to decentralisation strength, privacy preservation and suitability for constrained devices

| Framework Category | Performance Metrics | | |
|---|---|---|---|
| | Strength | Privacy | Constr. Devices |
| Federated Learning Based IDS/Anomaly Detection | High | High | Medium |
| Deep-learning-based trust/behavioural models | Medium | Medium | Low |
| Blockchain & Lightweight Ledger Framework | Very High | Medium | Medium |
| Zero Trust Oriented IOT Models | Medium | High | Medium |
| Decentralised Access Control & Identity | High | High | High |
| Graph/DRL/Hybrid Multi-layer Architectures | Medium | Medium | Low |

## 5.1. Performance Metrics: Accuracy and Detection Quality

Across the thirty reviewed works, most studies report very high detection or prediction accuracy, often above 98%, when evaluated on benchmark IoT datasets. However, the specific performance metrics, datasets and evaluation procedures vary notably between frameworks.

Deep-learning-based IDS and ML-driven anomaly detection systems consistently achieve near-perfect performance. For example, Douiba *et al.* [9], Hazman *et al.* [45] and Baksh *et al.* [44] report precision, recall and F1-scores close to 100% on datasets such as Bot-IoT, Edge-IIoTset, IoT-23 and CIC-IDS2017. El-Sofany *et al.* [21] similarly demonstrate that classical machine-learning classifiers, when combined with SDN/NFV-based orchestration, can maintain high detection rates in diverse IoT scenarios.

Federated deep-learning frameworks, such as SecFedDNN [6], BFLIDS [14] and the FL–DNN–blockchain model by Nazir *et al.* [35] achieved accuracy levels comparable to centralised training while preserving privacy and enabling distributed deployment. Ferrag *et al.* [20] provide detailed comparisons of centralised versus federated RNN, CNN and DNN models across Bot-IoT, MQTTset and TON_IoT, demonstrating that federated variants can match or exceed centralised accuracy when hyperparameters are tuned appropriately.

Works that apply FL outside of pure IDS contexts, such as Asha *et al.* [7] and Mohammed *et al.* [22] focus on prediction and optimisation tasks, including performance forecasting and secure transport scheduling, and similarly report high predictive accuracy.

Graph-based and Zero Trust frameworks, such as DFGL-LZTA by Zhou *et al.* [16] evaluate performance in terms of classification accuracy and improved Zero Trust policy decisions rather than focusing exclusively on intrusion detection. ZTA-IoT [17] emphasises policy correctness and enforcement consistency, using qualitative rather than dataset-driven metrics.

Baseline comparisons between centralised and decentralised architectures in Chen *et al.* [1], Chen *et al.* [2] and Khan *et al.* [3] further showed that decentralised approaches can maintain high accuracy even under data skew or cross-domain deployment, although they may introduce coordination overhead.

Overall, Table 2 and Table 3 indicate that federated-learning-based IDS and FL–blockchain hybrids score highly on performance, while survey-oriented works primarily contextualise these findings rather than reporting empirical accuracy.

## 5.2. Scalability and Communication Overhead

Scalability depends heavily on how architectures manage data exchange, model updates, consensus protocols and computation distribution.

Federated learning techniques improve scalability by transmitting model updates instead of raw data. SecFedDNN [6] and BFLIDS [14] employed gradient sparsification and update compression to reduce communication load. Manu *et al.* [13] explored asynchronous aggregation, demonstrating reduced waiting time and improved suitability for heterogeneous devices.

Multi-objective FL approaches, such as Mohammed *et al.* [22] explicitly balance energy, latency and accuracy through RL-driven client scheduling, showing that careful orchestration can maintain scalability without degrading learning performance.

Lightweight blockchain frameworks, such as COSIER [18] and GTxChain [8] addressed traditional blockchain limitations by introducing cluster-level chains, DAG-based validation, simplified consensus and off-chain storage, enabling reduced block-generation time and improved throughput.

Zero Trust and decentralised access-control architectures emphasise on policy scalability. BETAC-IoT [19] uses hybrid blockchains and Merkle trees to compactly maintain identity and authorisation information, whereas ZTA-IoT [17] uses layered policy enforcement to scale across large, heterogeneous IoT networks.

Survey works by Sah *et al.* [10], Diba *et al.* [23], Ferraris *et al.* [27] and Liu *et al.* [29], Dritsas et al. [30] consistently highlight FL and blockchain scalability challenges such as straggler issues, storage limits and communication overhead and recommend hierarchical designs and lightweight cryptography as practical solutions.

Overall, most of the reviewed frameworks achieved scalability either by partitioning computation across layers or by reducing data-transfer requirements, as reflected in the classification results in Table 2 and Table 3.

## 5.3. Privacy Preservation and Data Exposure

Privacy protection is a primary motivation for decentralised IoT architectures, and this is evident across the reviewed works.

Federated learning frameworks [5, 6, 13, 14, 22, 35, 38] enhance privacy by ensuring raw data remain local. DFGL-LZTA [16] expands this protection through local differential privacy and homomorphic encryption, preventing inference attacks during model aggregation. Ferrag *et al.* [20] and Jiang *et al.* [40] affirmed these privacy advantages; however, that gradient inference attacks remain as a matter of concern.

Blockchain-based privacy mechanisms appear in COSIER [18] and GTxChain [8] where ledger structures, off-chain storage and lightweight cryptography ensure tamper-evident tracking without exposing sensitive payloads. BETAC-IoT [19] and Bi *et al.* [5] used hybrid blockchains and Merkle trees to ensure integrity and confidentiality while supporting scalable identity management.

Hybrid cryptographic approaches, such as those in Sumathi *et al.* [12], combined classical cryptography with AI-based key management to protect data flows. Belchior *et al.* [26] highlighted the privacy implications of cross-chain interoperability, emphasising the need to control metadata leakage.

Zero Trust architectures such as ZTA-IoT [17], DFGL-LZTA [16] and the vehicle-focused ZT framework by Khan *et al.* [42] reinforced privacy by enforcing continuous verification, micro-segmentation and strict access-control boundaries.

As shown in Table 3, FL–blockchain hybrid systems achieved the strongest privacy scores, while centralised IDS approaches rated lower due to reliance on central data aggregation.

## 5.4. Trust Accuracy, Stability and Adaptiveness

Only a subset of the reviewed works maintains explicit trust scores; however, many frameworks implicitly provide trust signals through anomaly detection and behavioural classification.

Explicit trust-management frameworks include: Majdoubi *et al.* [4] who proposed decentralised trust establishment protocols resilient to collusion and Ferraris *et al.* [27] who analysed context-aware, multi-dimensional trust properties.

Behaviour-based and DNN-driven trust models such as Deep Trust [32] and NeuroTrust [33] demonstrate dynamic trust estimation and adaptive behaviour modelling in IoT environments. While these works emphasise trust computation rather than intrusion detection, they provide valuable insight into how deep learning can capture evolving device behaviour in decentralised systems.

FL-based IDS systems [5,6,14,22,35] often treat repeated malicious classification as implicit behavioural trust degradation.

Zero Trust and reinforcement-learning-driven models stand out for adaptiveness. DFGL-LZTA [16] uses a PPO agent to dynamically adjust security policies based on observed behaviour. BETAC-IoT [19] and Mohammed *et al.* [22] similarly update trust or access parameters in response to real-time conditions.

As reflected in Table 2, frameworks that incorporate persistent trust modelling or DRL-driven adaptation score higher for trust accuracy and adaptiveness than static or single-pass anomaly detectors.

## 5.5. Adversarial Robustness: Poisoning, Evasion and Identity Attacks

Adversarial robustness varies widely across the reviewed works. Poisoning resilience is addressed systematically in DFGL-LZTA by Zhou *et al.* [16] which uses reputation-filtered aggregation, DRL-guided defence, differential privacy and homomorphic encryption. Ferrag *et al.* [20] and Jiang *et al.* [40] highlighted poisoning risks in FL and recommended robust aggregation and anomaly scoring of updates.

Identity-based attacks are mitigated in blockchain-enabled systems such as BETAC-IoT [19], COSIER [18], and GTxChain [8] where cryptographic credentials and immutable logs reduce spoofing and replay risks. Liu *et al.* [29] showed how consensus and reputation mechanisms further mitigate Sybil attacks.

Evasion robustness is less consistently evaluated across the reviewed literature. Although many DL-based IDS claim high accuracy [9,44,45], few explicitly consider adversarial example generation or dynamic evasion strategies. Nicho *et al.* [24] further show that the computational cost of heavy CNN-based models can limit iterative retraining and adversarial hardening on resource-constrained IoT devices.

Table 2 reflects this imbalance: only a small number of frameworks demonstrate comprehensive robustness across poisoning, evasion and identity-level threats.

## 5.6. Computational and Energy Efficiency at the Edge

Practical deployment feasibility depends heavily on model size, computational cost and energy consumption.

Optimised ML and lightweight DNN designs, such as the Honey Badger–optimised RNN by Asha *et al.* [7] CatBoost by Douiba *et al.* [9] and PCC–XGBoost–LSTM by Hazman *et al.* [45] reduced computational cost and enabled near real-time inference. Nicho *et al.* [24] quantified the limitations of complex CNNs on microcontrollers, highlighting that accuracy gains may come at the expense of practical deployability on resource-constrained IoT and edge devices.

Lightweight blockchain designs such as COSIER [18] demonstrated reductions in memory, computation and storage requirements through ASCON cryptography and simplified consensus. Interoperability frameworks discussed by Belchior *et al.* [26] emphasised on the need for low-overhead cross-chain communication in IoT environments.

Resource-aware federated learning frameworks [6,13,22] employed sparsified aggregation, selective updates and asynchronous communication to improve efficiency. Surveys by Sah *et al.* [10] and Dritsas *et al.* [30] highlighted compression, quantisation and client selection as critical enablers for low-resource FL deployments.

As reflected in Table 3, lightweight blockchain and optimised ML approaches score highest for efficiency, while heavy DNNs and conceptual frameworks score lower.

## 5.7. Interoperability and Cross-Domain Applicability

Interoperability is addressed explicitly in only a limited number of the reviewed works but is increasingly recognised as essential for real-world deployment.

Blockchain interoperability is addressed in Belchior *et al.* [26] which outlined architectural requirements for cross-chain trust exchange. Liu *et al.* [29] highlighted the challenges in maintaining trust across multiple administrative domains.

FL-based IDS systems, such as Ferrag *et al.* [20], demonstrated cross-domain applicability by evaluating models across multiple datasets. Aldhaheri *et al.* [11] and Sah *et al.* [10] emphasised cross-domain variation in threat modelling and deployment constraints. Javed *et al.* [38] adapted FL–blockchain frameworks to vehicular IoT, highlighting suitability for mobile environments.

Multi-layer systems such as BETAC-IoT [19] and COSIER [18] are designed as domain-neutral trust and identity layers, with potential applicability in healthcare, transportation and smart-city environments. Mohammed *et al.* [22] focused on intelligent transport, while Asha *et al.* [7] developed architectures for specific domains but with generalisable components.

Table 2 and Table 3 indicate layered architectural frameworks score highest in terms of interoperability, while survey papers provide the broadest cross-domain coverage and conceptual interoperability analysis.

## 5.8. Strengths, Limitations and Design Trade-offs

A synthesis of the thirty articles covered in this study reveals several overarching patterns.

(i) **Strength**

- FL and DNN-based IDS provide high detection accuracy while enabling privacy-preserving and decentralised learning.
- Lightweight and hierarchical blockchains (e.g., COSIER) offer scalable, tamper-evident trust infrastructures.
- Zero Trust architectures formalise continuous verification and least privilege access in distributed environments.
- Surveys and conceptual analyses map the broader landscape and support the taxonomy developed in this review.

(ii) **Limitations**

- Many IDS focus on static datasets and lack evaluation under sophisticated adversarial conditions.
- Only a few FL frameworks explicitly mitigate poisoning and malicious-client attacks.
- Several architectures assume relatively powerful edge hardware, which may not reflect typical IoT constraints.
- Interoperability and standardised evaluation remain unresolved, despite frequent discussion in literature surveys.

(iii) **Design Trade-offs**

- **Accuracy vs. resource constraints:** Deep FL–DNN models deliver high accuracy but require more computation; lightweight ML scales better but may detect fewer sophisticated attacks.
- **Privacy vs. auditability:** Differential privacy and encryption improve confidentiality but add overhead, whereas blockchain increases transparency but may expose metadata.
- **Local autonomy vs. global consistency:** Decentralisation improves resilience but complicates synchronisation and uniform policy enforcement.

Overall, Tables 1–3 demonstrate that no single mechanism, such as FL, blockchain, ZTA or ML, fully addresses all IoT security challenges. Hybrid approaches that blend multiple decentralised mechanisms and adopt multi-layer architectures appear most effective, aligning with the goal of achieving secure, scalable and trust-aware IoT edge ecosystems.

## 6. Research Challenges and Future Directions

The comparative analysis, supported by Table 1, Table 2 and Table 3, shows that decentralised trust and security for IoT systems have advanced considerably, particularly through federated learning, lightweight blockchain architectures, Zero Trust enforcement, decentralised access control and hybrid graph- or DRL-enhanced models. Despite this progress, several critical challenges must be resolved before these techniques can be reliably deployed at scale across heterogeneous and resource-constrained edge environments.

The first challenge concerns robustness against advanced adversaries. Most intrusion detection and anomaly detection systems, including deep-learning-based approaches and federated IDS models, are evaluated chiefly in terms of accuracy and F1-score on the benchmark datasets, such as NSL-KDD, CIC-IDS2017, Bot-IoT and IoT-23. However, several studies, such as Ferrag *et al.* [20], Diba *et al.* [23] and Aldhaheri *et al.* [11], underline the vulnerability of FL and deep-learning pipelines to poisoning, evasion and backdoor attacks. Only a small number of frameworks integrate explicit defensive mechanisms. The DFGL-LZTA system by Zhou *et al.* [16] is a notable exception, combining federated graph learning with Zero Trust policies, local differential privacy, homomorphic encryption and DRL-based adaptive defence. Future research, therefore, must move beyond accuracy-driven evaluation and incorporate adversarial training, robust aggregation methods, gradient anomaly detection and formal defences against poisoning, evasion and identity manipulation.

The second major issue is the lightweight operation for constrained IoT hardware. Many effective approaches depend on deep neural networks, graph neural networks or reinforcement learning agents, all of which introduce computational and memory burdens. Studies, such as Nicho *et al.* [24] and Asha *et al.* [7], demonstrate that resource limitations often force computation to be shifted from sensors to fog or edge devices. Lightweight blockchain designs such as COSIER by Mershad [18], which employ ASCON-based cryptography, simplified consensus and hierarchical ledger structures, show how ledger mechanisms can be adapted to IoT constraints. Nevertheless, systematic methods for model compression such as pruning, quantisation, distillation and selective offloading across edge fog cloud tiers remain underdeveloped. Future work should prioritise ultra-lightweight FL and DNN architectures, energy-aware scheduling and dynamic model partitioning strategies that align security capabilities with the limited power, memory and processing resources of real-world IoT nodes.

The third challenge is interoperability across administrative domains, ledgers and trust models. Survey contributions by Belchior *et al.* [26], Liu *et al.* [29] and Ferraris *et al.* [27], highlight fragmentation in how trust is represented, exchanged and enforced. Current systems rely on incompatible combinations of FL-based anomaly scores, ledger-based credentials, reputation values, PKI identities and Zero Trust policy engines. Blockchain interoperability particularly cross-chain reputation or trust exchange remains emerging and is rarely addressed in practical deployments. Meanwhile, many IDS and trust-management models are trained on a single dataset or constrained to a specific domain (e.g., smart cities, industrial IoT, vehicular networks), limiting portability and generalisation. Future architectures should adopt standardised trust ontologies, cross-chain trust-token formats, domain-adaptive FL techniques and cross-dataset benchmarking practices to ensure interoperability across diverse IoT ecosystems.

The fourth challenge relates to balancing privacy and utility in decentralised learning. FL and blockchain assisted designs improve privacy by retaining raw data at the edge and securing model updates, as demonstrated in SecFedDNN [6], BFLIDS [14], BETAC-IoT [19], the frameworks by Nazir *et al.* [35] and Alzubi et al. [36], and domain-specific solutions by Mohammed *et al.* [22] and Javed *et al.* [38]. However, stronger privacy-preserving techniques, such as differential privacy, homomorphic encryption and secure multiparty computation, can degrade utility or introduce significant communication overhead. DFGL-LZTA illustrates that combining these techniques can be effective but requires careful tuning to avoid excessive accuracy loss or prohibitive costs. Future research should explore adaptive privacy mechanisms where protection levels adjust dynamically to threat conditions, alongside formal privacy–utility analyses that quantify trade-offs in realistic IoT environments.

The fifth challenge involves the representation, evolution and contextualisation of trust. Many reviewed systems generate binary labels or instantaneous anomaly scores without maintaining persistent, multi-dimensional trust states. While decentralised trust protocols, such as Majdoubi *et al.* [4] and identity schemes in BETAC-IoT [19], move toward explicit trust modelling, they remain limited in representing temporal evolution, context and cross-layer interactions. Future decentralised trust engines should integrate behavioural histories, identity attributes, environmental context (e.g., mobility, location, device type, etc.) and network topology, potentially using graph-based trust inference at the edge. Reinforcement learning agents, as used in Zhou *et al.* [16] and Mohammed *et al.* [22], could then update trust and risk estimates dynamically rather than relying on static thresholds.

The sixth challenge concerns the lack of real-world deployment and long-term evaluation. Many frameworks rely on offline datasets, simulators or idealised network assumptions. Only a few works report prototypes or extended deployment over real IoT environments. Operational constraints such as intermittent connectivity, node mobility, partial compromise, hardware degradation, or long-term drift in traffic patterns remain largely unaddressed. Future research should develop comprehensive edge testbeds integrating heterogeneous IoT devices with FL, blockchain and ZTA components, and evaluate systems under realistic conditions and adversarial threats. Longitudinal studies are also needed to track trust evolution, system stability and security performance over extended periods.

Finally, the most promising direction emerging from the reviewed literature is the convergence towards hybrid, multi-layer decentralised architectures. Frameworks such as COSIER, DFGL-LZTA, BETAC-IoT and a variety of FL–blockchain hybrids already demonstrated the potential of combining federated learning for distributed intelligence, lightweight blockchain for tamper-proof coordination, Zero Trust for continuous verification and decentralised identity management for fine-grained authorisation. Future IoT ecosystems are likely to rely on such integrated trust fabrics spanning edge, fog and cloud layers. Designing these fabrics to be interoperable, explainable, resource-aware and resilient while suitable for resource-constrained devices remains one of the most important open research challenges in the field.

# 7. Conclusion

This review article examined thirty recent studies, selected following the set inclusion and exclusion criteria, on decentralised trust and security mechanisms for IoT systems operating at the edge, focusing on federated learning, deep-learning-based detection, lightweight blockchain frameworks, Zero Trust architectures, decentralised identity and access control as well as hybrid models incorporating graph learning or reinforcement learning. Using a structured methodology and coherent thematic categorisation, the review outlined how IoT security is transitioning from traditional centralised architectures towards distributed, collaborative and context-aware designs.

Several clear trends emerged across the surveyed works. Federated learning has become a central technique for intrusion detection and behavioural modelling, enabling local training and privacy-preserving collaboration across heterogeneous nodes. Deep neural networks continue to provide strong detection accuracy, while lightweight blockchain and hybrid ledger architectures support scalable trust establishment, tamper-evident coordination and decentralised identity management. Zero Trust principles are increasingly embedded into edge systems to enforce continuous verification and least-privilege access, and hybrid frameworks combining graph learning, reinforcement learning and privacy-preserving mechanisms reflect the growing sophistication of decentralised security architectures.

The comparative analysis also highlighted several important limitations. Many current systems still lack resilience against poisoning, evasion and backdoor attacks, and resource constraints at the edge complicate deployment of computationally intensive models. Interoperability remains limited across heterogeneous IoT domains and privacy–utility trade-offs are not yet well quantified. Furthermore, trust mechanisms often lack persistent, context-aware modelling of device behaviour and real-world deployment remains rare, leaving open questions about scalability, robustness and long-term stability in operational environments. These gaps highlight the need for future research on lightweight adversarial defences, cross-domain trust representation, adaptive privacy controls and practical hybrid architectures suited to real-world constraints.

Overall, the findings indicate that decentralised trust and security mechanisms hold considerable promise for enabling resilient, privacy-preserving and scalable protection in modern IoT ecosystems. Realising this potential will require continued research that integrates robust adversarial defences, efficient distributed learning, interoperable trust fabrics and comprehensive real-world evaluation. The convergence of federated learning, blockchain, Zero Trust Architecture and behavioural analytics suggests a clear trajectory towards fully decentralised, multi-layered IoT security architectures capable of supporting the complexity, scale and heterogeneity of next-generation cyber-physical systems.

## Acknowledgements

This research is financially supported by Xiamen University Malaysia (Project codes: XMUMRF/2021-C8/IECE/0025 and XMUMRF/2022-C10/IECE/0043].